\documentclass[nofootinbib]{revtex4}
\usepackage{graphicx}
\usepackage{amsmath,amssymb,epsfig}
\usepackage{paralist}
\usepackage{comment}
\usepackage{wrapfig}

\renewcommand{\vec}[1]{\boldsymbol{\mathrm{#1}}}

\begin{document}

\title{Search for gravitationally lensed interstellar transmissions}

\author{Slava G. Turyshev}

\affiliation{
Jet Propulsion Laboratory, California Institute of Technology,\\
4800 Oak Grove Drive, Pasadena, CA 91109-0899, USA}

\date{\today}

\begin{abstract}
   
We explore interstellar light transmission facilitated by gravitational lensing, focusing on axially-symmetric lensing configurations where the transmitter, lens, and receiver are nearly aligned. Positioning an optical transmitter in the lens's focal region, we investigate the caustic formed by a diffraction-limited annular beam of light emitted by the transmitter. We analyze the impact of the lens's point spread function (PSF) on the projected beam's structure, estimate the power delivered to a receiver at interstellar distances, and assess the major noise sources. We determine detection sensitivity in both noise- and signal-dominated regimes. Considering realistic assumptions about the transmitter's performance, we explore signal detection strategies enhanced by the spatial broadening of the received beam, a result of the transmitting lens's PSF. Our findings indicate that detecting lensed optical signals from nearby stars is achievable using established optical engineering technologies. A network of spatially distributed astronomical facilities capable of observations in multiple narrow spectral bands will enhance the search. Our results support the feasibility of interstellar power transmission via gravitational lensing, directly contributing to ongoing optical SETI efforts.
  
\end{abstract}


\maketitle 

\section{Introduction}

Interstellar power transmission is very challenging. Even for a collimated laser beam, the large distances involved result in a very small energy received. Consider, a diffraction-limited telescope with aperture $d_{\tt T}$ yielding the beam divergence of $\theta_0\simeq\beta(\lambda/d_{\tt T})=1.00 \times 10^{-6}\,(\lambda/1\,\mu{\rm m})(1\,{\rm m}/d_{\tt T})$, where $\beta$ is a coefficient
that depends on the type of light amplitude distribution and the definition of beam diameter, typically of $\beta\simeq {\cal O}(1)$. When the signal reaches the receiver at distance $z$, the beam is expanded to a spot with the radius of $\rho_*\simeq z(\lambda/d_{\tt T})\simeq 3.09 \times10^{11}\,{\rm m}~(z/ 10\,{\rm pc})(\lambda/1\,\mu{\rm m})(1\,{\rm m}/d_{\tt T})$. As a result, in the case of a free-space laser power transmission in the vacuum, a telescope with the aperture $d_{\tt R}$ receives only a small fraction of the transmit power $P_{\tt T}$ given as $P_{\tt R}= P_{\tt T}  { \pi ({\textstyle\frac{1}{2}}d_{\tt R})^2}/{\pi \rho_*^2}$  or 
 {}
\begin{eqnarray}
P_{\tt R}&=& P_{\tt T}\frac{ \pi ({\textstyle\frac{1}{2}}d_{\tt R})^2}{\pi z^2} \Big(\frac{d_{\tt T}}{\lambda}\Big)^2 
\simeq 
2.63 \times 10^{-24} \Big(\frac{P_{\tt T}}{1~{\rm W}}\Big)\Big(\frac{1~\mu{\rm m}}{\lambda}\Big)^2
\Big(\frac{d_{\tt T}}{1~{\rm m}}\Big)^2
\Big(\frac{d_{\tt R}}{1~{\rm m}}\Big)^2\Big(\frac{10~{\rm pc}}{z}\Big)^2~ {\rm W},~~
\label{eq:po-free}
\end{eqnarray}
which yields photon flux of only $Q_{\tt R}=(\lambda/h c)P_{\tt R}\simeq1.32 \times 10^{-5}$\, phot/s for $\lambda=1\,\mu$m.
These signals are weak and require both 1) precise and  stable transmitter pointing toward us to within $\rho_{\tt L}/z\simeq 
{\rm few}\,\mu$rad and 2) us having instruments capable of detecting such signals against the challenging optical backgrounds that are present in our stellar neighborhood.  
Therefore, unless there is a significant effort to send us a message using a powerful, focused beam directly aimed at us, detecting faint and transient signals with optical SETI\footnote{See details on the ongoing SETI efforts at \url{https://en.wikipedia.org/wiki/Search_for_extraterrestrial_intelligence}} remains extremely challenging.

The situation changes if a transmitter is placed in the focal region of a stellar gravitational lens to benefit from the lens' significant light amplification. Lensing geometries where the light source, the lens, and the receiver are in an approximate alignment maximize efficiency of the interstellar communication links  \cite{Turyshev:2024}. Given our current  technological maturity, we cannot yet position a transmitter in the focal region of the solar gravitational lens. However, we now begun to understand the physics of the interstellar power transmission facilitated by gravitational lensing and are in a position to devise practical strategies to search for the signals originated around nearby stars. These efforts can rely on the already existing technologies, bringing this topic in the realm of advanced photonics and optical engineering.

This paper is organized as follows: In Section~\ref{sec:wave}, we present the wave-theoretical tools to describe light propagation in a gravity field. We discuss the caustic formed by the diffraction-limited beam of light. We consider diffraction of light in a spherically-symmetric gravitational field and consider the process of the interstellar power transmission. In Section \ref{sec:SNR-sec}, we estimate the levels of the anticipated signals, introduce major noise sources, and evaluate signal-to-noise ratio (SNR). In Section~\ref{sec:detect}, we discuss various detection strategies. Our conclusions are presented in Section~\ref{sec:summary}.

\section{EM waves in a gravitational field}
\label{sec:wave}

\begin{figure}[t!]
  \begin{center}
\includegraphics[width=0.70\linewidth]{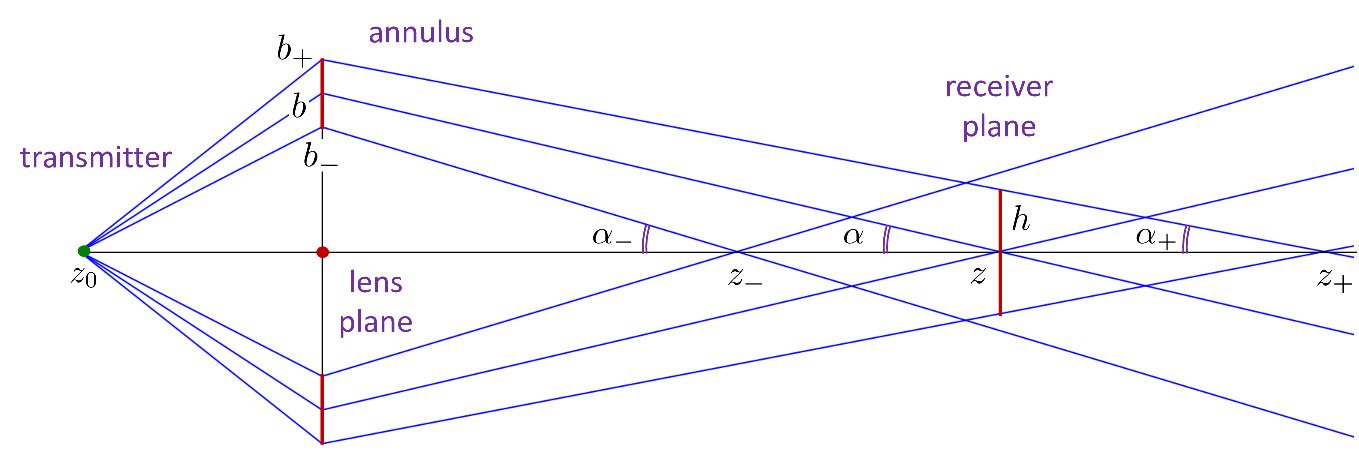}
  \end{center}
  \vspace{-20pt}
  \caption{Annular beam propagation in a thin lens approximation showing transmitter, lens,  annulus, and receiver plane.}
\label{fig:geom-beam}
\end{figure}

We consider a nearby star with mass $M_{\tt L}$, radius $R_{\tt L}$, and its Schwarzschild radius of $r_g=2GM_{\tt L}/c^2$.  We assume  a transmitter placed on the  optical axis—the line connecting the solar system and this stellar gravitational lens, as illustrated in Fig.\,\ref{fig:geom-beam}. We also assume that this transmitter is facing the lens and is positioned in the lens' focal region at a distance $z_0 > R^2_{\tt L}/2r_{g}=547.8\, (R_{\tt L}/R_\odot)^2(M_\odot/M_{\tt L})$\,AU from it \cite{Lodge:1919,Chwolson:1924,Einstein:1936,vonEshleman:1979,Turyshev-Andersson:2002,Turyshev-Toth:2017}, on the side opposite the solar system. Next, we assume that this is a CW laser optical transmitter that is capable of coherently illuminating an annulus around the lens with a radius of $b=\sqrt{2r_g z_0}> R_{\tt L}$, which is chosen to ensure the beam's partial focusing at the receiver \cite{Turyshev:2024}. To achieve this, a diffraction-limited transmitter must produce an annular beam of light.\footnote{See \cite{Duocastella-Arnold:2012,Livneh-etal:2018} on the current efforts in producing annular laser beams.} This beam should have the mean diverging angle of $b/z_0 \simeq 7.16 \times 10^{-6}\,({b}/{R_{\tt L}})(R_{\tt L}/{R_\odot})({650\,{\rm AU}}/{z_0})~ {\rm rad}$ relative to the optical axis. As shown by Fig. \ref{fig:geom-beam}, there is a finite-size caustic that is formed in the focal region of the lens. What determines the dimensions of this caustic? In the following discussion, we examine the geometry of the region where beam focusing occurs.

\subsection{Lensing caustic}
\label{sec:caustic}

It is known that when a plane EM wave, originating at infinity, travels near a massive body, its wavefront experiences bending (see Figs.~2, 3 in \cite{Turyshev-Toth:2017}). In the weak field approximation of general relativity, this deflection angle is $\theta_{\tt gr}={2r_g}/{b}$, where $b$ is a light rays's impact parameter. As a result, a massive celestial body acts as a lens by focusing the EM radiation (i.e., where the light rays with the same $b$  intersect the optical axis) at distance $z$, that is determined as
{}
\begin{eqnarray}
\frac{b}{z}=\theta_{\tt gr}
\qquad \Rightarrow \qquad z= \frac{b^2}{2r_g}
\qquad {\rm or} \qquad
b=\sqrt{2r_g z}. 
\label{eq:foc_z}
\end{eqnarray}

Clearly, the rays with different impact parameters will intersect the optical axis at different distances, which is consistent with lens' spherical aberration. As a result, for an infinitely large incident plane wave coming from infinity, a monopole gravitational lens forms a caustic with a semi-infinite focal line that begins from $z_0=R_{\tt L}/2r_g$.   

In our case, the situation is reversed: the transmitter, placed in the lens's focal region at a large but finite distance $z_0 > R_{\text{L}}/2r_g$ from it, faces the lens and emits a finite-width beam of light towards it. What shape and dimensions will the caustic, forming in the focal region behind the lens, have?

To answer this question, we will use the geometrical optics, the small angles, and the thin lens approximations. When the light rays are coming from a source that is located at a large but finite distance from the lens, $z_0$, for them to intersect the lens plane with the same impact parameter $b$, they must be emitted at an angle $\beta=b/z_{0}$. As the rays travel through the thin lens plane, direction of their travel is affected by lens' gravity. As a result, the angle at which the rays with the same $b$ will intersect the optical axis now is smaller and  is given as $\alpha=\theta_{\tt gr}-b/z_{0}$, which has to be positive (see \cite{Turyshev:2024} for discussion of the impact of different values of $\alpha$ on focusing), yielding
{}
\begin{eqnarray}
\frac{b}{z}= \frac{2r_g}{b}-\frac{b}{z_0}
\qquad \Rightarrow \qquad b= \sqrt{2r_g \tilde z},
\qquad {\rm where}
\qquad \tilde z =\frac{z_0z}{z_0+z}. 
\label{eq:tilde=}
\end{eqnarray}
This expression implies that if the transmission distance, $z_0$, and the reception distance, $z$, are known, then the transmitter has to the lens plane illuminate at a particular impact parameter $b= \sqrt{2r_g \tilde z}$, as determined by (\ref{eq:tilde=}). 

With $\alpha < \theta_{\text{gr}}$, the distance $z = b / \alpha$, at which light rays with the same impact parameter $b$ intersect the optical axis (see Fig.\,\ref{fig:geom-beam}), increases compared to the case of a plane wave originating from infinity, which was described by (\ref{eq:foc_z}). In \cite{Turyshev-Toth:2019-extend,Turyshev:2024}, we have shown that the expression for $z$ given (\ref{eq:foc_z}) is now modified to read
{}
\begin{eqnarray}
\frac{b}{z}= \frac{2r_g}{b}\Big(1-\frac{b^2}{2r_gz_0}\Big) \equiv \alpha 
\qquad \Rightarrow \qquad z= \frac{b^2}{2r_g}\frac{1}{1-b^2/2r_gz_0} . 
\label{eq:foc_z2}
\end{eqnarray}

We now consider a diffraction-limited beam of light leaving the  telescope at a divergence angle of $\sim\lambda/d_{\tt T}$, where $d_{\tt T}$ is the aperture of a telescope used to transmit the signal. When the beam reaches the lens plane, it produces an annulus with the mean radius of $b$ and the thickness of $2\Delta b$ with $\Delta b=(\lambda/d_{\tt T})z_0\simeq 9.72\times 10^7 (\lambda/1\,\mu{\rm m})(1\,{\rm m}/d_{\tt T})(z_0/650\,{\rm AU})\,{\rm m}\simeq 0.14 R_\odot$. The entire transmitted energy is deposited into the annulus with the area of  $A=4\pi b\Delta b=4\pi b(\lambda/d_{\tt T})z_0$. As the beam propagates further, it is affected by the lens' spherical aberration. 

By the time the beam reaches the focal region, located at an interstellar distance, the beam has expanded, forming a caustic, see Fig.\,\ref{fig:geom-beam}.  The caustic features an axially symmetric shape resembling a bicone, consisting of two conjoined cones that meet at their bases yet have slightly different heights. What are the diameter and the length of this caustic? 
 
To investigate the dimensions of this caustic, we will use the thin lens approximation. Fig.\,\ref{fig:geom-beam} shows the beam with the mean impact parameter $b$. The divergence of this beam, due to optical diffraction, produces an annulus on the lens plane where the two most distant rays have impact parameters $b_{\pm} = b \pm \Delta b$. 

Similar to (\ref{eq:foc_z2}), we can determine the angles $\alpha_{\pm}$ and the distances $z_{\pm}$ at which these rays intersect the optical axis:
{}
\begin{eqnarray}
\frac{b_\pm}{z_\pm}=\frac{2r_g}{b_\pm}\Big(1-\frac{b^2_\pm}{2r_gz_0}\Big)\equiv\alpha_\pm
\qquad \Rightarrow \qquad 
z_\pm= \frac{b^2_\pm}{2r_g}\frac{1}{1-b^2_\pm/2r_gz_0} . 
\label{eq:foc_z2-pm}
\end{eqnarray}

Taking into account that $\Delta b\ll b$, indeed
{}
\begin{eqnarray}
\frac{\Delta b}{b}=\frac{(\lambda/d_{\tt T})z_0}{b}= 0.14
\Big(\frac{\lambda}{1\,\mu{\rm m}}\Big)\Big(\frac{1\,{\rm m}}{d_{\tt T}}\Big)\Big(\frac{z_0}{650\,{\rm AU}}\Big)\Big(\frac{R_{\tt L}}{b}\Big)\Big(\frac{R_\odot}{R_{\tt L}}\Big),
\label{eq:db-b}
\end{eqnarray}
we may develop all the relevant expressions to the first order in $\Delta b/b$. 

First of all, we compute the width of the beam at the point $z$, where the central ray  intersects the optical axis. As seen in Fig.\,\ref{fig:geom-beam}, the diameter of this area is given by $d_{z}=2h=2\alpha_+(z_+-z)$. Using (\ref{eq:foc_z2}) and (\ref{eq:foc_z2-pm}), we express this as 
{}
\begin{eqnarray}
d_{z}=2h=2b_+\Big(1-\frac{z}{z_+}\Big)=\frac{4\Delta b}{1-b^2/2r_g z_0}+{\cal O}\big(\Delta b^2\big).
\label{eq:width}
\end{eqnarray}

Similarly, we determine the length of the caustic $\Delta z=z_+-z_-$ as it extends alone the optical axis
{}
\begin{eqnarray}
\Delta z=z_+-z_- =\frac{4\Delta b}{1-b^2/2r_g z_0}\frac{z}{b}+{\cal O}\big(\Delta b^2\big).
\label{eq:length}
\end{eqnarray}
With result for the impact parameter $b=\sqrt{2r_g\tilde z}$ from (\ref{eq:tilde=}), we express $1-b^2/2r_g z_0=1-\tilde z/z_0=z_0/(z_0+z)$. Using this result in (\ref{eq:width})--(\ref{eq:length}), and considering $\Delta b = (\lambda/d_{\text{t}})z_0$, we determine the diameter and length of the caustic as  below
{}
\begin{eqnarray}
d_z&=&\frac{4\lambda}{d_{\tt T}}(z_0+z)=8.25
\Big(\frac{\lambda}{1\,\mu{\rm m}}\Big)\Big(\frac{1\,{\rm m}}{d_{\tt T}}\Big)\Big(\frac{z}{10\,{\rm pc}}\Big)~{\rm AU},
\label{eq:width}\\
\Delta z&=& \frac{zd_z}{\sqrt{2r_g\tilde z}}=
\frac{d_z}{\theta_{\tt ER}}
=16.29
\Big(\frac{\lambda}{1\,\mu{\rm m}}\Big)\Big(\frac{1\,{\rm m}}{d_{\tt T}}\Big)\Big(\frac{z}{10\,{\rm pc}}\Big)^2\Big(\frac{M_\odot}{M_{\tt L}}\Big)^\frac{1}{2}\Big(\frac{650\,{\rm AU}}{z_0}\Big)^\frac{1}{2}~{\rm kpc},
\label{eq:length}
\end{eqnarray}
where $\theta_{\tt ER}$ is the radius of the Einstein ring as seen by the observer at $z$ (see (\ref{eq:theta1-t}) below for details).

Interestingly, the dimensions of the caustic created by a monopole lens for an annular beam are finite but very large. The diameter $d_z$ is a factor 2 larger than that for a free-space propagation. This is due beam broadening by the monopole gravitational lens. From a practical standpoint, increasing the transmitting telescope's diameter would result in a more compact caustic, thereby enhancing the power density at the receiver thus aiding in signal detection. 

Although light is present everywhere within the bicone-shaped caustic areas with dimensions of $(\Delta z, d_z)$ given by (\ref{eq:width})--(\ref{eq:length}), not all regions within it are of interest to us. In fact, as shown in Fig.\,\ref{fig:geom-beam}, similarly to the scenario with a monopole lens and an incident plane wave \cite{Turyshev-Toth:2017}, the propagation of an annular diffraction-limited beam gives rise to four distinct regions: shadow, geometrical optics, and both weak and strong interference regions, namely:
\begin{inparaenum}[i)]
\item Since there is no light with impact parameters that fall outside the range estublished by the diffraction limit, $b\in[b_-,b_+]$, the corresponding areas form a shadow region behind the lens, where no light from the transmitter is present;
\item The regions through which only one ray from the source passes at each point  (i.e., outside the interference region set by the biconical shape in Fig.\,\ref{fig:geom-beam}) correspond to the regions of geometric optics;
\item As we enter the bicone-shaped area, we move into the zone of weak interference, where at least two rays from opposite sides of the annulus intersect. Here, the light intensity begins to increase, marking it as the region of weak interference;
\item  As we approach closer to the optical axis,  in the immediate vicinity of it,  at separations  $r_{\tt GL}\leq r_g (z/z_0)\simeq 9.37 \times 10^6(M_{\tt L}/M_\odot)(z/10\,{\rm pc})(650\,{\rm AU}/z_0)$\,m, we enter the region of strong interference. Here, near the optical axis, an intensely concentrated beam is present. 
\end{inparaenum}
Proper description of the EM field in the interference regions requires a wave-theoretical treatment, which we discuss next.

\subsection{Light amplification}
\label{sec:power-X-sig}

At a high level, we know that when electromagnetic (EM) waves  travel in the vicinity of a stellar gravitational lens, its gravity causes the waves to scatter and diffract \cite{Turyshev:2017,Turyshev-Toth:2017}.  Due to the spherical aberration inherent in a monopole lens, the beam will not focus at a single point. Instead, the lens's gravitational field will cause the light to converge within the finite-size caustic forming the focal region in the vicinity of the optical axis, as discussed in \cite{Turyshev:2024} and shown in Fig. \ref{fig:geom-beam}.  The properties of this caustic are driven by the transmission annulus, with the mean radius of $b$ and the thickness of $2\Delta b \sim2(\lambda/d_{\tt T})z_0$.  According to (\ref{eq:tilde=}), for a properly chosen transmitter position, $z_0$, the transmitted light will be deposited in the solar system at distance  $z\gg z_0$, where the beam will be compressed and partially focused.  A solar system observer,  looking back at the lensing star, will see the transmitted signal in the form of the Einstein ring around the lens. Compared to a free-space propagation (\ref{eq:po-free}), such gravitational focusing results in a major increase of the power density of the EM field deposited in the receiver plane \cite{Turyshev:2024}.  Here, we consider the physics of this process. 

\begin{figure}
  \begin{center}
\includegraphics[width=0.55\linewidth]{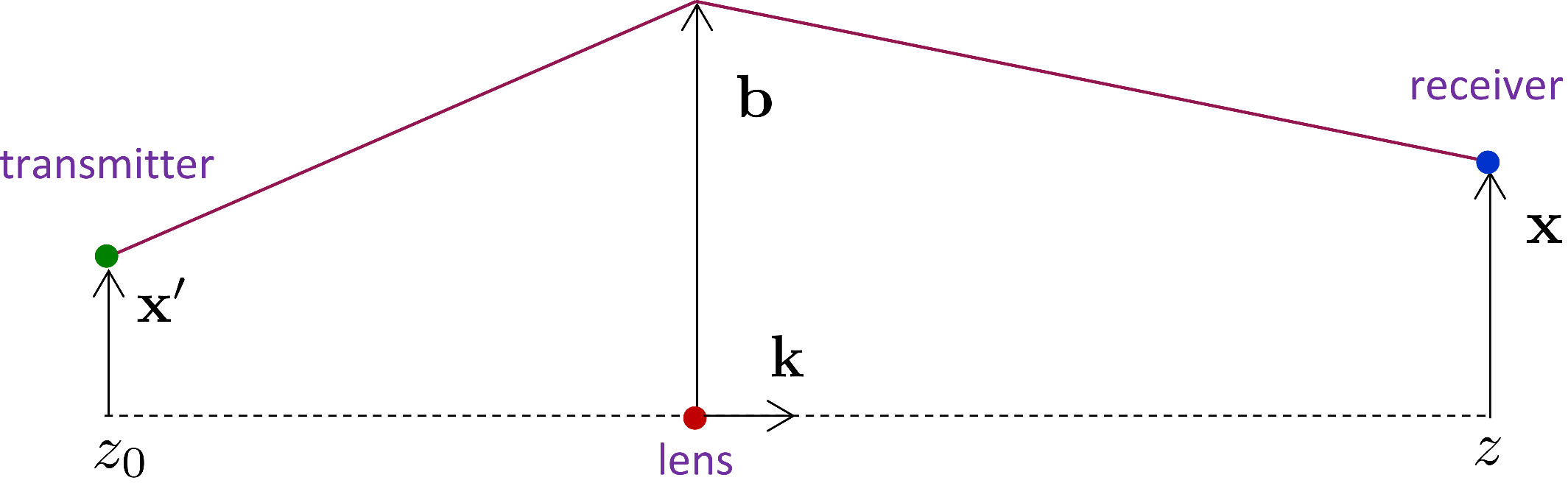}
  \end{center}
  \vspace{-20pt}
  \caption{A lens-centric  geometry for interstellar power transmission via gravitational lensing showing transmitter,  lens, and  receiver. Also shown is the distance from the lens to the transmitter plane, $z_0$, and that from the lens to the receiver plane, $z$.  
  }
\label{fig:geom}
\end{figure}

We introduce a lens-centric cylindrical coordinate system  $(\rho,\phi,z)$ with its  $z$-axis oriented along the unperturbed direction of the incident wave's propagation (see Fig.~\ref{fig:geom}), given by a unit vector $\vec k$. We also introduce a light ray's impact parameter, $\vec b$, with $\vec b \perp \vec k$. In this coordinate system, a ray of light is emitted from a point with coordinates $(\vec x',-z_0)$, while it is received in the receiver plane at a point with coordinates $(\vec x,z)$ and 
{}
\begin{align}
{\vec x}'=\rho'(\cos\phi',\sin \phi',0), \qquad
{\vec b}=b(\cos\phi_\xi,\sin \phi_\xi,0), \qquad 
{\vec x}=\rho(\cos\phi,\sin \phi,0).
\label{eq:note-x}
\end{align}

Assuming validity of the eikonal and the thin lens approximations, and considering a uniform surface brightness of the source, the Fresnel-Kirchhoff diffraction formula \cite{Landau-Lifshitz:1988,Born-Wolf:1999} yields the following expression for the wave's amplitude at the observer (receiver) location
 {}
\begin{eqnarray}
A(\vec x', \vec x) &=&E_0
 \frac{k}{i z_0z}\frac{1}{2\pi}\iint d^2\vec b e^{ik S(\vec x', \vec b, \vec x)},
  \label{eq:amp-FK}
\end{eqnarray}
where $k=2\pi/\lambda$ is the wavenumber and  $S(\vec x', \vec b, \vec x)$  is the effective path length (eikonal) along the path from the source at $(\vec x', -z_0)$ to the observer's position at $(\vec x,z)$ via a point $(\vec b,0)$ on the lens plane  (see Fig.~\ref{fig:geom} and
\cite{Turyshev:2024})
 {}
\begin{eqnarray}
S(\vec x', \vec b, \vec x) &=&
 \sqrt{(\vec b-\vec x')^2+z_0^2}+ \sqrt{(\vec b-\vec x)^2+z^2}+\psi(\vec b)= \nonumber\\
 &=&z_0+z+\frac{(\vec x-\vec x')^2}{2(z_0+z)}+\frac{z_0+z}{2z_0z}\Big(\vec b-\frac{z_0}{z_0+z}\Big(\vec x+\frac{z}{z_0}\vec x'\Big)\Big)^2+\psi(\vec b) + {\cal O}(b^4),
  \label{eq:amp-S}
\end{eqnarray}
where $\psi(\vec b) $ is the gravitational phase shift that is acquired by the wave as it moves along its geodetic path from the source to the observer.  For a monopole lens it has the form   \cite{Turyshev-Toth:2021-multipoles,Turyshev-Toth:2022-STF}:
{}
\begin{eqnarray}
\psi(\vec b)&=& \frac{2}{c^2}\int_{z_0}^zdzU(\vec b, z)=kr_g\ln 4k^2zz_0-2r_g\ln k b.
  \label{eq:gp}
\end{eqnarray}

As a result, the wave amplitude on the receiver plane (\ref{eq:amp-FK}) can be written as
{}
\begin{eqnarray}
A(\vec x',\vec x)&=& \frac{E_0}{z_0+z}\exp\Big[ik\Big(z_0+z+\frac{(\vec x-\vec x')^2}{2(z_0+z)}\Big)\Big]\cdot F(\vec x',\vec x),
  \label{eq:amp00}
\end{eqnarray}
where $F(\vec x)$ is the amplification factor that is given by the following diffraction integral \cite{Turyshev-Toth:2021-multipoles,Turyshev:2024}
{}
\begin{eqnarray}
F(\vec x', \vec x) &=&
 \frac{ke^{ikr_g\ln 4k^2zz_0}}{i \tilde z}\frac{1}{2\pi}
 \iint d^2\vec b\exp\Big[ik\Big(
 \frac{1}{2\tilde z}\Big(\vec b-\frac{\tilde z}{z}\Big(\vec x+\frac{z}{z_0}\vec x'\Big)\Big)^2-2r_{g}\ln k b\Big)\Big],
  \qquad  
 \tilde z=\frac{z_0z}{z_0+z}.~~~~
  \label{eq:amp-A1*}
\end{eqnarray}
Note that the phase factor under the integral is the Fermat potential along the light path.

Benefiting from the axial-symmetry, we integrate over $\phi_\xi$ from 0 to $2\pi$. Then, to integrate over the impact parameter, $b$, we use the method of stationary phase \cite{Turyshev-Toth:2017} and present (\ref{eq:amp-A1*}) as below (see details in \cite{Turyshev:2024})
{}
\begin{eqnarray}
F(\vec x',\vec x) &=&
\sqrt{2\pi kr_{g}}  e^{i\varphi}
J_0\Big(k\frac{\sqrt{2r_{g} \tilde z}}{z} \rho(\vec x',\vec x) \Big),
\qquad ~~
\rho(\vec x',\vec x)=\big|\vec x+\frac{z}{z_0}\vec x'\big|,
  \label{eq:amp-A1*24}
\end{eqnarray}
where  the impact parameter for which the phase is stationery was determined to be $b=\sqrt{2r_{g} \tilde z}$. Also, $\varphi=\sigma_0+kr_g\ln 2k(z_0+z)$ with $\sigma_0=-kr_g\ln (kr_g/e)-\frac{\pi}{4}$, see  \cite{Turyshev-Toth:2017}, and  where $J_0$ is the Bessel function of the first kind \cite{Abramovitz-Stegun:1965}.

\begin{figure}
\begin{minipage}[b]{.46\linewidth}
\rotatebox{90}{\hskip 14pt  {\footnotesize Amplification ($\times 10^{10}$)}}
\includegraphics[width=0.8\linewidth]{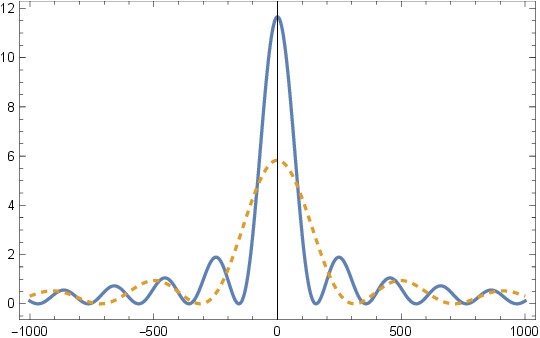}
\vskip -5pt
\rotatebox{0}{\hskip 25pt  {\footnotesize Distance from the optical axis, $\rho$ [m]}}
\end{minipage}
\begin{minipage}[b]{.46\linewidth}
\includegraphics[width=0.73\linewidth]{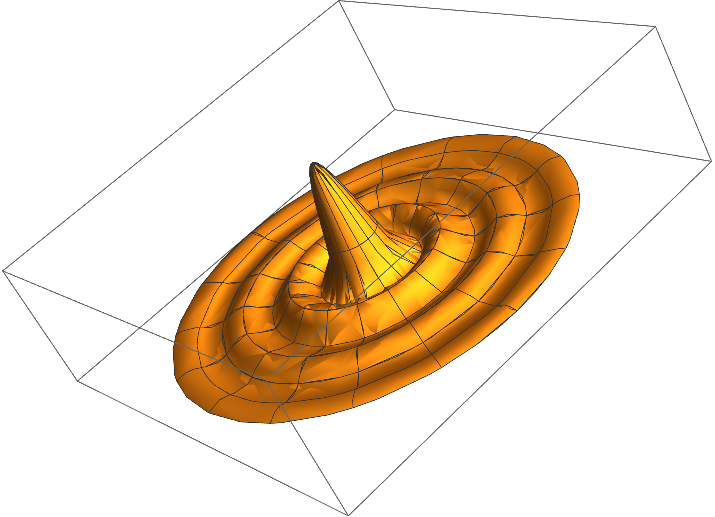}
\end{minipage}
  \vspace{-10pt}
  \caption{Left: amplification and PSF pattern of the transmitting lens (\ref{eq:S_z*6z-mu2}) projected on the solar system plotted for two wavelengths. The solid line represents   $\lambda=1.0\,\mu{\rm m}$, the dotted line is for $\lambda=2.0\,\mu{\rm m}$. Right: a three-dimensional representation of the PSF in the image plane for $\lambda=1.0\,\mu$m with the peak corresponding to direction along the transmitter-lens line. }
\label{fig:psf}
  \vspace{-5pt}
\end{figure}

To consider the light amplification, we determine the point-spread function (PSF), which, in this case, is just the square of the amplification factor (\ref{eq:amp-A1*24}) \cite{Turyshev-Toth:2017}, namely
{}
\begin{align}
{\tt PSF}(\vec x',\vec x)&
=2\pi k r_{g}  J^2_0\Big(k\frac{\sqrt{2r_{g} \tilde z}}{z}
\rho(\vec x',\vec x) \Big) \simeq 1.17\times 10^{11} J^2_0\Big(k\frac{\sqrt{2r_{g} \tilde z}}{z} \rho(\vec x',\vec x)\Big)\, \Big(\frac{M_{\tt L}}{M_\odot}\Big)\Big(\frac{1\,\mu{\rm m}}{\lambda}\Big).
\label{eq:S_z*6z-mu2}
\end{align}
Fig.~\ref{fig:psf} shows the resulting Airy pattern formed by the lens in an image plane in the solar system. With this result, we establish the angular resolution of the transmitting lens which is determined by the first zero of the Bessel function $J_0(x)$ in (\ref{eq:S_z*6z-mu2})  \cite{Turyshev-Toth:2020-extend}, which occurs at $x=2.40483$, yielding the angular resolution measured in nano-arcseconds (nas):
{}
\begin{eqnarray}
r_{\tt GL}=\big|\frac{{\vec x}}{ z}+\frac{{\vec x'}}{{ z}_0}\big|=\frac{2.40483}{k\sqrt{2r_g\tilde z}}=0.38 \frac{\lambda}{\sqrt{2r_g\tilde z}}\simeq 0.10\Big(\frac{\lambda}{1\,\mu{\rm m}}\Big)\Big(\frac{650\,{\rm AU}}{ z_0}\Big)^\frac{1}{2}~{\rm nas}.
\label{eq:S_=}
\end{eqnarray}
We observe that the form of resolution $r_{\tt GL}$ is similar to that of a classical diffraction-limited telescope that behaves as $\sim1.22\,\lambda/d$. However, in the case of a gravitational lens, the denominator is  the impact parameter $\sqrt{2r_g\tilde z}=b\gtrsim R_{\tt L}$, which is why result (\ref{eq:S_=}) yields a very high resolution behaving as $r_{\tt GL}\lesssim 0.38\, \lambda/R_{\tt L}$. 
This underscores the fact that after passing by the lens and traveling a large distance from it, the beam becomes tightly compressed.

An observer, positioned on the optical axis at $z \gg z_0$ from the gravitational lens, will see an Einstein ring around the  lens with the radius $\theta_{\tt ER}$ given as:\footnote{Note that in typical microlensing scenarios, where the source is at much larger distance $z_0$ from the lens then an observer is, $z$, the following is valid: $z_0\gg z$. In this case, the size of the Einstein ring is, $
\theta_{\tt ER}={\sqrt{2r_{g} \tilde z}/{z}}\simeq\sqrt{{2r_g}/{z}}=1.38 \times 10^{-7}\,
\sqrt{M_{\tt L}/M_\odot}\,\sqrt{10~{\rm pc}/ z}$ or $\sim {\cal O}(28.54~{\rm mas})$, which is much larger than that estimated in (\ref{eq:theta1-t}) and could be resolved by the largest instruments available \cite{Turyshev-Toth:2023}.}
{}
\begin{eqnarray}
\theta_{\tt ER}=\frac{\sqrt{2r_{g} \tilde z}}{z}\simeq \frac{\sqrt{2r_{g}  z_0}}{z}
 = 2.46 \times 10^{-9}\,\Big(\frac{M_{\tt L}}{M_\odot}\Big)^\frac{1}{2}\Big(\frac{z_0}{650\,{\rm AU}}\Big)^\frac{1}{2}\Big(\frac{10\,{\rm pc}}{z}\Big)~ {\rm rad},
  \label{eq:theta1-t}
\end{eqnarray}
or $\sim {\cal O}(0.5~{\rm mas})$, which is quite challenging to resolve with  the current generation of optical instruments (see \cite{Turyshev-Toth:2023}). 

We observe that the first zero of the projected PSF pattern  (\ref{eq:S_z*6z-mu2}) occurs at the distance of 
{}
\begin{eqnarray}
\rho_{\tt GL}=\frac{2.40483}{k\theta_{\tt ER}}\simeq 155.84\,{\rm m}\,\Big(\frac{\lambda}{1~\mu{\rm m}}\Big)\Big(\frac{M_\odot}{M_{\tt L}}\Big)^\frac{1}{2}\Big(\frac{650~{\rm AU}}{z_0}\Big)^\frac{1}{2}\Big(\frac{z}{10\,{\rm pc}}\Big),
  \label{eq:rho-t}
\end{eqnarray}
from the optical axis, which is much larger than any modern optical telescope \cite{Turyshev-Toth:2023}. (Note that this PSF broadening  is consistent with the effect of a lens with spherical aberration.)
Thus, (\ref{eq:S_z*6z-mu2}) is good for very small separations, $\rho\simeq \rho_{\tt GL}$. However, in realistic cases, assuming that the signal is pointed at the Sun,  receiver's misalignment may be as much as $\rho \sim \,1\,{\rm AU}\gg \rho_{\tt GL}$. In this case, we use the approximation for the Bessel functions for large arguments \cite{Abramovitz-Stegun:1965,Turyshev-Toth:2019-blur}:
{}
\begin{eqnarray}
J^2_0\Big(k\frac{\sqrt{2r_{g} \tilde z}}{ z}\rho\Big)\simeq \frac{1+\sin\big(2k\theta_{\tt ER}\rho\big)}{\pi k\theta_{\tt ER}\,\rho}\simeq \frac{20.59\,{\rm m}}{\rho}
\Big(\frac{\lambda}{1\,\mu{\rm m}}\Big)
\Big(\frac{M_\odot}{M_{\tt L}}\Big)^\frac{1}{2}\Big(\frac{650\,{\rm AU}}{z_0}\Big)^\frac{1}{2}\Big(\frac{z}{10\,{\rm pc}}\Big),
\label{eq:BF}
\end{eqnarray}
were  we used the fact that for large arguments $\rho\gg \rho_{\tt GL}$ the function $\sin(2k\theta_{\tt ER}\rho)$  is rapidly oscillating; thus, it averages to 0. Using this in (\ref{eq:S_z*6z-mu2}),  the PSF for $\rho\gg \rho_{\tt GL}$ takes the form
{}
\begin{align}
{\tt PSF}(\rho)& \simeq 2.40\times 10^{12} \Big(\frac{1\,{\rm m}}{\rho}\Big) \Big(\frac{M_{\tt L}}{M_\odot}\Big)^\frac{1}{2}\Big(\frac{650\,{\rm AU}}{z_0}\Big)^\frac{1}{2}\Big(\frac{z}{10\,{\rm pc}}\Big).
\label{eq:S_z*6z-mu2+}
\end{align}

As a result, the spatial structure of the transmitting lens's PSF, as depicted in Fig.\,\ref{fig:psf} and described by (\ref{eq:S_z*6z-mu2}), will influence the received signal. This structure is well-characterized: at typical wavelengths, its spatial frequency in the image plane scales to hundreds of meters, as specified by (\ref{eq:theta1-t}), and its amplitude is inversely proportional to the radial distance from the center of the pattern, as illustrated by (\ref{eq:S_z*6z-mu2+}) and (\ref{eq:powed-l-Q}). These characteristics of the PSF's spatial distribution (\ref{eq:S_z*6z-mu2}) and (\ref{eq:S_z*6z-mu2+}) will be encoded in the structure of the received signal, thus directing our search strategy.

\subsection{Power  
projected  by a stellar gravitational lens}
\label{sec:image-photom}
\label{sec:SGL-imaging-ext}

Examining the PSF (\ref{eq:S_z*6z-mu2}) and recognizing that the angular resolution (\ref{eq:S_=}) is extremely small, we see that a monopole gravitational lens acts as a convex lens by focusing light but also by expanding the projected beam, according to ${\vec x}=-({ z}/{z_0}){\vec x}'$. This mapping rule suggests that by the time the transmitted beam reaches the solar system, its size will have increased by a factor of ${ z}/{z_0}\sim3.17\times 10^{3}\,({ z}/10\,{\rm pc}) (650\,{\rm AU}/z_0)$. As a result, the diameter of the transmitting telescope as seen by an observer  (i.e., the brightest spot) will be expanded to fill in a larger area of
{}
\begin{equation}
d_{\tt beam}=\frac{ z}{z_0}d_{\tt T}\simeq3.17\,\Big(\frac{z}{10\,{\rm pc}}\Big) \Big(\frac{650\,{\rm AU}}{z_0}\Big)\Big(\frac{d_{\tt T}}{1~{\rm m}}\Big)~{\rm km}.
\label{eq:rE}
\end{equation}
This is the diameter of the  aperture of the transmitting telescope projected onto the image plane by the PSF of a monopole gravitational lens (\ref{eq:S_z*6z-mu2}). It does not account for the diffraction of the transmitted beam; thus, it is much smaller compared to the diameter of the entire caustic given by  (\ref{eq:width}). However, it gives the size of the beam with the highest intensity characteristic to the strong interference region.  

Even the largest telescopes today will not be able to see the entire beam at once. (We dealt with a similar situation while considering the imaging of exoplanets with the SGL \cite{Turyshev-Toth:2019-blur}.) Using (\ref{eq:rE}), we see that a telescope with the aperture $d_{\tt R}\sim 10$\,m could resolve the projected laser beam with diameter of $d_{\tt beam}$ with $N$ linear resolution elements given by
{}
\begin{equation}
N=\frac{d_{\tt beam}}{d_{\tt R}}=\frac{ z}{z_0}\frac{d_{\tt T}}{d_{\tt R}}\simeq 317.33\,\Big(\frac{z}{10\,{\rm pc}}\Big) \Big(\frac{650\,{\rm AU}}{z_0}\Big)\Big(\frac{d_{\tt T}}{1\,{\rm m}}\Big)\Big(\frac{10\,{\rm m}}{d_{\tt R}}\Big).
\label{eq:res-el}
\end{equation}

This behavior is consistent with the PSF broadening, as given by (\ref{eq:S_z*6z-mu2}) and (\ref{eq:rho-t}), and its impact on the expansion of a light beam emitted from the focal point of a convex lens exhibiting spherical aberration. With this insight, we need to treat a laser transmitter operating in the focal region of a gravitational lens as an extended source. 

Therefore, we consider the laser transmitter to be an extended luminous source with surface brightness $B(\vec x')$ with dimensions of ${\rm W \,m}^{-2}{\rm sr}^{-1}$, the power density, $I_0(\vec x)$, received in the receiver plane is computed by integrating the PSF (\ref{eq:S_z*6z-mu2}) over the surface of the extended source and is given as below  (with $\theta_{\tt ER}$ from (\ref{eq:theta1-t}))
{}
\begin{eqnarray}
I_0(\vec x)&=&\frac{2\pi k r_{g} }{(z_0+z)^2}\iint d^2\vec x'\, B(\vec x')\,
J^2_0\Big(k\theta_{\tt ER}
\big|{\vec x}+\frac{ z}{z_0}{\vec x}'\big|\Big).
\label{eq:power_dens}
\end{eqnarray}

A telescope with aperture $d_{\tt R}\ll d_{\tt sig}$, centered at a particular point $\vec{x}_0$ in the image plane, will receive the signal $P(\vec x_0)=\iint d^2\vec x \,I_0\big(\vec x_0+\vec x\big),$ where the integration is done within the telescope's aperture $|\vec x|\leq {\footnotesize\frac{1}{2}}d_{\tt R}$, and with $|{\vec x}_0+{\vec x}|\leq {\footnotesize\frac{1}{2}}d_{\tt beam}$. This yields a result that depends on the telescope's position on the receiver plane:
{}
\begin{eqnarray}
P_{\tt R}({\vec x}_0)&=&\frac{2\pi k r_{g} }{(z_0+z)^2}
\hskip -3pt\iint\displaylimits_{|{\vec x}|^2\leq (\frac{1}{2}d_{\tt R})^2}\hskip -3pt d^2\vec x 
\hskip -5pt \iint\displaylimits_{|{\vec x'}|^2\leq (\frac{1}{2}d_{\tt T})^2}\hskip -5pt d^2\vec x' B(\vec x')\,
J^2_0\Big(k\theta_{\tt ER}
\big|{\vec x}_0+{\vec x}+\frac{ z}{z_0}{\vec x}'\big|\Big).
\label{eq:power_rec2}
\end{eqnarray}

In the case of a laser for which the emission is highly collimated which may conservatively be characterized by the beam divergence of $\theta_0=\beta(\lambda/d_{\tt T})$, with $\beta\simeq {\cal O}(1)$. As a result, the laser luminosity is scaled as $L_0= P_{\tt T}/\theta_0^2$. Although laser beams exhibit Gaussian spatial profile, for simplicity, we will assume a top-hat profile with uniform brightness, so, that the source's surface brightness takes the form: $B(\vec x')=P_{\tt T}/\theta_0^2 \pi (\frac{1}{2}d_{\tt T})^2$. Using this expression in (\ref{eq:power_rec2}), yields
{}
\begin{eqnarray}
P_{\tt R}({\vec x}_0)&=&\frac{2\pi k r_{g} }{(z_0+z)^2}\frac{P_{\tt T}}{\pi (\frac{1}{2}d_{\tt T})^2}\Big(\frac{d_{\tt T}}{\lambda}\Big)^2
\hskip -3pt\iint\displaylimits_{|{\vec x}|^2\leq (\frac{1}{2}d_{\tt R})^2}\hskip -3pt d^2\vec x 
\hskip -5pt \iint\displaylimits_{|{\vec x'}|^2\leq (\frac{1}{2}d_{\tt T})^2}\hskip -5pt d^2\vec x' \,
J^2_0\Big(k\theta_{\tt ER}
\big|{\vec x}_0+{\vec x}+\frac{ z}{z_0}{\vec x}'\big|\Big),
\label{eq:power_rec2s}
\end{eqnarray}
which, in the case when gravity is absent, is identical to (\ref{eq:po-free}). Otherwise, the difference is due to the PSF presence.  

To evaluate this integral, we use the same approach developed in \cite{Turyshev-Toth:2019-blur,Turyshev-Toth:2020-extend}. As a result, defining $\rho_0=|\vec x_0|$, the power received at the detector placed in the focal plane of an optical telescope is given as 
{}
\begin{eqnarray}
P_{\tt R}(\rho_0)&=&P_{\tt T}\frac{\pi ({\textstyle\frac{1}{2}}d_{\tt R})^2}{\pi (z_0+z)^2}\Big(\frac{d_{\tt T}}{\lambda}\Big)^2\,\frac{4\sqrt{2r_g \tilde z}}{d_{\tt T}}\,\mu(\rho_0),
\qquad{\rm with}\qquad
\mu(r_0)=\frac{2}{\pi}{\tt Re} \Big\{ {\tt E}\Big[\Big(\frac{\rho_0}{{\textstyle\frac{1}{2}}d_{\tt beam}}\Big)^2\Big]\Big\},
\label{eq:power_rfin}
\end{eqnarray}
where,  following  \citep{Turyshev-Toth:2020-extend,Turyshev-Toth:2019-blur,Turyshev-Toth:2019-image}, we introduced the blur factor $\mu(\rho_0)$ due to the spherical aberration of the monopole lens with ${\tt E}[x]$ being the elliptic integral \citep{Abramovitz-Stegun:1965}  and ${\tt Re}\{\}$ is the operation taking the real part of a complex quantity.

The behavior of $\mu(\rho_0)$  is shown in Fig.~6 of \cite{Turyshev-Toth:2020-extend}. This factor captures blur's contribution present on the image plane due to  spherical aberration of a monopole lens. It reaches its maximal value of 1 for receiver on the optical axis, i.e., $\mu(0)=1$. On the edge of the projected beam, it decreases to  $\mu({\textstyle\frac{1}{2}}d_{\tt beam})=0.64$. Outside the footprint of the beam projected on the solar system, $\mu(\rho_0)$  falls off as $\propto {\textstyle\frac{1}{2}}d_{\tt beam}/\rho_0$, as expected from the PSF of a monopole lens \cite{Turyshev-Toth:2022a}.

Comparing the results for a free-space propagation, given by  (\ref{eq:po-free}), and that for gravitationally-lensed scenarios, from  (\ref{eq:power_rfin}), we see that treating transmitter as an extended and resolved source, the light amplification is given as  
{}
\begin{eqnarray}
{\cal A}_{\tt GL}(\rho_0)&=&\frac{4\sqrt{2r_g \tilde z}}{d_{\tt T}}\,\mu(\rho_0)=3.03\times10^9\, \mu(\rho_0) \,
\Big(\frac{M_{\tt L}}{M_\odot}\Big)^\frac{1}{2}
\Big(\frac{z_0}{650\,{\rm AU}}\Big)^\frac{1}{2}\Big(\frac{1~{\rm m}}{d_{\tt T}}\Big),
\label{eq:ampl}
\end{eqnarray}
which is less compared to the amplification for point sources, given by (\ref{eq:S_z*6z-mu2}), as expected, but still very high. 

Eq.~(\ref{eq:power_rfin}) describes the power of the light received from the transmitted laser beam that is present in the receiver plane, both inside and outside the beam's footprint.  Such behavior of the signal is due to the specific optical properties of the lens given by its PSF (\ref{eq:S_z*6z-mu2}) which, as a function of the distance to the optical axis on the image plane, falls out much more slowly than the PSF of a regular telescope (see discussion in \cite{Turyshev-Toth:2019-blur}). This fact provides valuable insight for the search of a transmitted signal  and the relevant work on prospective search campaign planning and development.

\section{Detection sensitivity} 
\label{sec:SNR-sec}

Understanding the detection sensitivity is crucial in the search for transmitted laser signals. This is done by evaluating the signal-to-noise ratio (SNR) in realistic transmission scenarios, which will be done below.

\subsection{Evaluating anticipated signals}
\label{sec:power-X}

To assess the effectiveness of the interstellar power transmission with a stellar gravitational lens, we consider the same energy transmission scenario used to derive (\ref{eq:po-free}). 
We again assume that transmission is characterized by the beam divergence set by the telescope's aperture $d_{\tt T}$ yielding angular resolution of $\theta_0\simeq\lambda/d_{\tt T}=1.00 \times10^{-6}~(\lambda/1\,\mu{\rm m})(1\,{\rm m}/d_{\tt T})\,{\rm rad}$. When the signal reaches the receiver at the distance of $(z_{\tt t}+z_{\tt r})$ from the transmitter, the beam is expanded to a large spot with the radius of $\rho_*=(z_{\tt t}+z_{\tt r})(\lambda/d_{\tt T})$. However, in this case, we need account for the fact that while passing by the lens, the light intensity is amplified according to (\ref{eq:S_z*6z-mu2}). This is due to the fact that the lens collimates the otherwise diverging beam of light, thus delivering a larger power density to the receiver (see Fig.\,\ref{fig:geom-beam}). As a result, based ion (\ref{eq:power_rfin}) a receiver telescope gets a larger fraction of the transmitted power  
 {}
\begin{eqnarray}
P_{\tt R}^{\tt GL}&=&P_{\tt T}\frac{\pi ({\textstyle\frac{1}{2}}d_{\tt R})^2}{\pi (z_0+z)^2}\Big(\frac{d_{\tt T}}{\lambda}\Big)^2\,\frac{4\sqrt{2r_g z_0}}{d_{\tt T}}\,\mu(\rho_0) 
\simeq\nonumber\\
&\simeq&
 7.95 \times 10^{-15}\,\mu(\rho_0) \Big(\frac{P_{\tt T}}{1~{\rm W}}\Big)\Big(\frac{1~\mu{\rm m}}{\lambda}\Big)^2 \Big(\frac{d_{\tt T}}{1~{\rm m}}\Big)
\Big(\frac{d_{\tt R}}{1~{\rm m}}\Big)^2\Big(\frac{10~{\rm pc}}{z}\Big)^2\Big(\frac{M_{\tt L}}{M_\odot}\Big)^\frac{1}{2}\Big(\frac{z_0}{650\,{\rm AU}}\Big)^\frac{1}{2}~{\rm W}.~
\label{eq:powed-l}
\end{eqnarray}  

Comparing  (\ref{eq:powed-l})  and (\ref{eq:po-free}), we see that power transmission link aided by a gravitational lens amplifies the received power by $P^{\tt GL}_{\tt R}/P^{\tt free}_{\tt R} \simeq 3.03\times10^9\,\mu(\rho_0)$, as given by (\ref{eq:ampl}).  

To estimate photon flux at the receiver, we compute  $Q_{\tt GL}=(\lambda/hc)P^{\tt GL}_{\tt R}$ from (\ref{eq:powed-l}), yielding
{}
\begin{eqnarray}
\hskip -2pt
Q_{\tt GL}(\rho) \hskip -2pt &\simeq& \hskip -2pt 
4.01 \times 10^{4}\,\mu(\rho_0) \Big(\frac{P_{\tt T}}{1~{\rm W}}\Big)\Big(\frac{1~\mu{\rm m}}{\lambda}\Big)^2 \Big(\frac{d_{\tt T}}{1~{\rm m}}\Big)
\Big(\frac{d_{\tt R}}{1~{\rm m}}\Big)^2\Big(\frac{10~{\rm pc}}{z}\Big)^2\Big(\frac{M_{\tt L}}{M_\odot}\Big)^\frac{1}{2}\Big(\frac{z_0}{650\,{\rm AU}}\Big)^\frac{1}{2}~~ {\rm phot/s}.
\label{eq:powed-l-Q}
\end{eqnarray}

Thus, due to gravitational lensing, the receiver experiences a significant photon flux. To assess the viability of this transmission link, it's essential to evaluate how major noise sources affect the overall detection sensitivity.

\subsection{Major noise source} 
\label{sec:noise}

The angular radius of a stellar lens is $R_{\tt L}/z=2.26\times 10^{-9}(R_{\tt L}/R_\odot)(10\,{\rm pc}/z)\,{\rm rad}$, which is similar to that of the size of the  Einstein ring formed by the transmitted beam of light (\ref{eq:theta1-t}). As the angular resolution of the largest  telescopes today (see \cite{Turyshev-Toth:2023}) is $\lambda/d_{\tt R}\simeq 1\times 10^{-7}(\lambda/1\,\mu{\rm m})(10\,{\rm m}/d_{\tt T})$,  they will not be able to resolve neither the star nor the ring, rendering the use of a coronagraph impractical. Consequently, the brightness of the lensing star, as captured by the telescope, constitutes the primary source of noise that needs to be mitigated. Other potential sources of noise, such as the stellar background, the star's atmosphere, and the interstellar medium, could be considered for a more detailed analysis in future stages, if necessary. However, they are not expected to provide significant contributions.  

To estimate the relevant photon flux, we consider our Sun. When dealing with laser light propagating in its vicinity, we need to be concern with the flux within some bandwidth $\Delta \lambda$ around the laser wavelength $\lambda$, assuming we can filter the light that falls outside $\Delta \lambda$. Taking the Sun's temperature to be $T_\odot=5\,772$~K, we estimate the solar brightness from the Planck's radiation law and derive the luminosity of the Sun within a narrow bandwidth:
{}
\begin{align}
L_\odot (\lambda, \Delta \lambda)
&{}= 4\pi^2 R^2_\odot
 \frac{2hc^2 }{\lambda^5\big(e^{hc/\lambda k_B T_\odot}-1\big)}\Delta \lambda
\simeq 2.05 \times 10^{24} \Big(\frac{1\,\mu{\rm m}}{\lambda}\Big)^5\Big(\frac{\Delta \lambda}{10\,{\rm nm}}\Big)~~   {\rm W},
\label{eq:model-L0*t}
\end{align}
where  $\sigma$ and $k_B$ are the Stefan-Boltzmann and the Boltzmann constants. For a given lens' brightness, $L_{\tt L}$, we use (\ref{eq:model-L0*t}),  to derive the photon flux from the lensing star received by a telescope that is given by $Q_{\tt \star} = ({\lambda}/{hc})L_{\tt L} (\lambda, \Delta \lambda)  {\pi ({\textstyle\frac{1}{2}}d_{\tt R})^2}/{\pi z^2} $, yeilding:
{}
\begin{eqnarray}
Q_\star &=&
2.71\times 10^7\,\Big(\frac{L_{\tt {\tt L}}}{L_\odot}\Big)\Big(\frac{1\,\mu{\rm m}}{\lambda}\Big)^4\Big(\frac{\Delta \lambda}{10\,{\rm nm}}\Big)
\Big(\frac{d_{\tt R}}{1~{\rm m}}\Big)^2\Big(\frac{10~{\rm pc}}{z}\Big)^2~~~{\rm phot/s}.
\label{eq:pow-fp==2}
\end{eqnarray}

Thus, despite the strength of the stellar flux (\ref{eq:model-L0*t}), its impact can be significantly mitigated by  using a narrow bandpass filter at the receiver, as shown in (\ref{eq:pow-fp==2}). Note that there is already available technology that can be used to search for CW laser signals —specifically, spectrographs with a resolution $R=\lambda/\Delta\lambda\sim10^5$ within the 1–5 $\mu$m bandwidth \cite{Kaeufl-etal:2004}, which suggests a $\Delta \lambda = 0.05$ nm. Clearly, utilizing this specific $\Delta \lambda$ value can further reduce the stellar background noise contribution (\ref{eq:pow-fp==2}) and enhance detection sensitivity, as detailed by \cite{Howard-etal:2004}. However, the application of such a narrow bandwidth complicates the search effort for yet unknown signal by presuming prior knowledge of the transmitting wavelength $\lambda$. Consequently, below we will use a broader bandpass filter value used in (\ref{eq:pow-fp==2}).   

\subsection{Signal to noise ratio} 
\label{sec:SNR}

Quantitative assessment of the SNR, is a key to identifying and characterizing received signals. We use results (\ref{eq:powed-l-Q}) and (\ref{eq:pow-fp==2}) and estimate the  SNR as usual
\begin{align}
{\rm SNR}=\frac{Q_{\tt GL}}{\sqrt{Q_{\tt GL}+Q_{\tt \star}}}.
\label{eq:snr}
\end{align}

Below, we will consider the detection sensitivity in the noise- and  signal-dominated regimes.  As our objective here is to evaluate the feasibility of interstellar power transmissions facilitated by gravitational lensing, to this end, we will rely on straightforward estimates based on the relevant physical parameters, omitting the engineering aspects of the hardware necessary for establishing the laser link (for relevant discussions see \cite{Horwath:1996,Howard-etal:2004,Clark:2018}.)

When noise dominates the signal, $Q_{\tt \star}\gg Q_{\tt GL}$, the SNR is given as  
{}
\begin{align}
{\rm SNR}_{\tt nd}(\rho)&=\frac{Q_{\tt ER}}{\sqrt{Q_{\tt \star}}}{}\simeq 7.69\,\mu(\rho_0)\Big(\frac{P_{\tt T}}{1~{\rm W}}\Big) \Big(\frac{L_\odot}{L_{\tt {\tt L}}}\Big)^\frac{1}{2}\Big(\frac{10\,{\rm nm}}{\Delta \lambda}\Big)^\frac{1}{2}\Big(\frac{d_{\tt T}}{1~{\rm m}}\Big) \Big(\frac{d_{\tt R}}{1~{\rm m}}\Big)\Big(\frac{10~{\rm pc}}{z}\Big)\Big(\frac{M_{\tt L}}{M_\odot}\Big)^\frac{1}{2}\Big(\frac{z_0}{650\,{\rm AU}}\Big)^\frac{1}{2}\,\sqrt{\frac{t}{1\,{\rm s}}},
\label{eq:snr-cor-t}
\end{align} 
where we used conservative assumptions on the set of parameters characterizing the interstellar transmission link.  Below, we will explore  the relevant parameter space that may be lead to improvements in the SNR estimate (\ref{eq:snr-cor-t}).  

We note that the estimate (\ref{eq:snr-cor-t}) was derived assuming a modest transmission power of 1 W. However, this power could potentially be much higher in practical applications. For instance, modern commercially available industrial lasers already feature powers of 1--50 kW. Additionally, the technique of coherent beam combination—merging multiple transmitting laser beams—could elevate the total transmitted power to several gigawatts (GW) or more. Furthermore, both the transmitting and receiving apertures could be significantly larger than those previously considered. Consequently, the strength of the received signal could vastly surpass that of stellar noise, to the extent of completely overshadowing the star. This scenario necessitates a different approximation method when evaluating (\ref{eq:snr}).

Therefore, in the case, when the signal dominates the noise, $Q_{\tt GL}\gg Q_{\tt L}$, 
  (\ref{eq:snr}) yields
  {}
\begin{eqnarray}
{\rm SNR}_{\tt sd}(\rho)=\sqrt{Q_{\tt GL}}&\simeq&
6.33\times 10^{3}\sqrt{\mu(\rho_0)}
\Big(\frac{P_{\tt T}}{1~{\rm kW}}\Big)^\frac{1}{2}\Big(\frac{1~\mu{\rm m}}{\lambda}\Big) \Big(\frac{d_{\tt T}}{1~{\rm m}}\Big)^\frac{1}{2}
\Big(\frac{d_{\tt R}}{1~{\rm m}}\Big)\Big(\frac{10~{\rm pc}}{z}\Big)\Big(\frac{M_{\tt L}}{M_\odot}\Big)^\frac{1}{4}\Big(\frac{z_0}{650\,{\rm AU}}\Big)^\frac{1}{4}\,\sqrt{\frac{t}{1\,{\rm s}}},~~~~
\label{eq:powed-l-Qs}
\end{eqnarray}
which is, again, quite conservative, but useful for our purposes. Similarly to  (\ref{eq:snr-cor-t}), result (\ref{eq:powed-l-Qs}) shows the SNR on the optical axis with factor $\mu(\rho_0)$ modulating the signal for any potential misalignments.  

We note that although the estimates  (\ref{eq:snr-cor-t}) and (\ref{eq:powed-l-Qs}) were developed under modest assumptions on  the parameters of a transmission link, they  indicate high levels of detection sensitivity. These results support the feasibility of projecting significant power across interstellar distances with the assistance of gravitational lensing.

\section{Detection strategies} 
\label{sec:detect}

Detecting laser signals presents several challenges. The first hurdle is the highly monochromatic nature of laser light, which emits at a specific wavelength. This characteristic necessitates pinpointing the exact wavelength, $\lambda$, to match an extraterrestrial signal and selecting a suitable narrow band-pass filter width, $\Delta \lambda$, to enhance the SNR, particularly in noise-dominated environments (\ref{eq:snr-cor-t}). Another challenge is the directionality of laser transmissions. Unlike radio waves, which can spread in all directions, laser beams are highly focused and narrow. Thus, for interstellar laser signals to be detected, they must be precisely aimed towards the Earth, especially since the interstellar gas and dust, though largely transparent to near-IR light, do not alleviate the need for exact alignment. Consequently, while the stellar brightness (\ref{eq:pow-fp==2}) and optical background noise are relatively well understood for each star in our stellar neighborhood, the specific link parameters of the transmitted signal (\ref{eq:powed-l-Q})—such as the transmitted power, aperture, wavelength, and pointing direction used for that purpose—are not known in advance.

Gravitational lensing makes some challenges above to be less critical, but some will require consideration: Although the beam of light we are searching for is significantly brighter compared to that of a free-space laser transmission, the transmitted beam arrives in the solar system being highly compressed. Clearly, this feature must be accounted for and be appropriately addressed at the transmitter, thus to enhance the probability of signal detection.
 
 \subsection{Considering a search campaign}
\label{sec:weak-int2}

 For interstellar transmissions utilizing gravitational lensing, positioning a transmitter within the focal region of a stellar lens is essential. Assuming the requisite technology for such a deployment exists, it is reasonable to also assume the transmitter would have adequate propulsion capabilities. Consequently, we can infer that the transmitter is designed to counteract the relative proper motion between the lens and the Sun, accurately directing its signal towards the Sun while compensating for the necessary stellar dynamics. This approach, as discussed in \cite{Turyshev-Toth:2022-wobbles} within the context of exoplanet imaging via the solar gravitational lens (SGL), underscores the importance of integrating advanced propulsion and precise navigational strategies in the transmitter's design.
 
As discussed in Sec.~\ref{sec:caustic}, the caustic formed by the annular beam's propagation around the lens is large. According to (\ref{eq:width}), its diameter can extend across several AUs, namely $d_z\simeq 8.25({\lambda}/{1\,\mu{\rm m}})({1\,{\rm m}}/{d_{\tt T}})({z}/{10\,{\rm pc}})\,{\rm AU}$. Moreover, within the strong interference region, where the intensity of received light peaks, the distance from the optical axis may be as large as $r_{\tt GL}\leq r_g (z/z_0)\simeq 9.37 \times 10^6(M_{\tt L}/M_\odot)(z/10\,{\rm pc})(650\,{\rm AU}/z_0)\,{\rm m}\simeq 1.51 R_\oplus$. 
The most intense light, however, is observed when the transmitting telescope aligns directly with the receiver, yielding a received beam diameter notably smaller than in other regions, as discussed in Sec.~\ref{sec:image-photom} and given by (\ref{eq:rE}): $d_{\tt beam}\simeq3.17\,({z}/{10\,{\rm pc}}) ({650\,{\rm AU}}/{z_0})({d_{\tt T}}/{1~{\rm m}})~{\rm km}$. These are the three scales where we need to search for the signal.

Therefore, we find that in optical communications utilizing gravitational lenses, precise aiming of the signal transmissions is also crucial. There could be multiple strategies for initiating transmission. For instance, in one scenario, the transmission could be so precisely directed that Earth passes through the targeted spot. Consequently, it’s reasonable to assume that the transmitter would have the capability to track Earth's movement. Given this precision, one might question whether a deliberately wider beam, capable of encompassing the entire Earth, would be employed instead. This is just few of many scenarios that merit thorough exploration.

As an example, we consider a scenario in which a transmitter targets a specific spot 1 AU from the Sun, allowing Earth to intersect this spot annually. The observability of this lensing event depends on the diameter of the received beam and Earth’s orbital velocity. In a signal-dominated scenario, the SNR, as shown in (\ref{eq:powed-l-Qs}), is influenced by the factor $\sqrt{\mu(\rho_0)}$, which outside the projected beam diminishes as $({\textstyle\frac{1}{2}}d_{\tt beam}/\rho_0)^\frac{1}{2}$, decreasing slowly with distance from the optical axis. Consequently, even covering the entire Earth, with $\rho_0\simeq R_\oplus$, the SNR remains exceptionally high at $\sim 6.33\times 10^{3}({\textstyle\frac{1}{2}}d_{\tt beam}/R_\oplus)^\frac{1}{2}\simeq 99.79$, ensuring detectability. Moreover, even at lunar distances, the SNR, estimated at around $12.76$, is sufficiently strong for detection, indicating a robust signal across significant distances.

To model the dynamics of the event, we assume a vanishing impact parameter. Under such conditions, the duration of a lensing event, $ \Delta t_{\tt beam}$, or the time for the receiver to cross the entire diameter of the strong interference region  is  
{}
\begin{eqnarray}
\Delta t_{\tt beam}&=& 
\frac{2r_{\tt GL}}{v_\oplus}
\simeq 1.06
\Big(\frac{M_{\tt L}}{M_\odot}\Big)\Big(\frac{ z}{10~{\rm pc}}\Big)\Big(\frac{650~{\rm AU}}{ z_0}\Big)\Big(\frac{30~{\rm km/s}}{v_\oplus}\Big)~~{\rm min}.
\label{eq:n-lens-dura}
\end{eqnarray}

Depending on the SNR required for signal detection (\ref{eq:powed-l-Qs}), this estimate may be improved. As the radius of the search region increases as $\rho^{\tt SNR}_0\sim (6.33\times 10^{3}/{\rm SNR})^2{\textstyle\frac{1}{2}}d_{\tt beam}\simeq 6.35\times 10^8\,(10/{\rm SNR})^2 \,{\rm m}$,  for ${\rm SNR}=10$ it reaches $\simeq 99.57\, R_\oplus$, or nearly 1.7 times of the lunar distance.  In this case, the estimate (\ref{eq:n-lens-dura}) increases to $\Delta t_{\tt beam}\simeq 5.88$~hr. Clearly, as we move away from the optical axis while assuming $P_{\tt T}=1$\,kW as in (\ref{eq:powed-l-Qs}),  the photon flux (\ref{eq:powed-l-Q}) drops as $\sim 4.01\times 10^7 ({\textstyle\frac{1}{2}}d_{\tt beam}/\rho_0)\, {\rm phot/s}$, which at $\rho_0=99.57\, R_\oplus$, reaches the value of $\sim 1.1\times 10^2\,{\rm phot/s}$, all received within the narrow bandwidth of $\Delta \lambda=10$\,nm as in (\ref{eq:pow-fp==2}). This is detectable with the current generation of optical instruments. 

We note that increasing the receiver's aperture to $d_{\tt R}=10$\,m  in the example above not only increases the flux by a factor of 100 but also improves the SNR by a factor of 10, thereby significantly enhancing the chances of detection. In fact, there is a rather significant parameter space that may be used to develop a number of practical search strategies. Thus, given that the actual transmitter link parameters could exceed those assumed in (\ref{eq:powed-l-Qs}), the peak SNR could be significantly higher. Key variables such as the transmit power ($P_{\tt T}$) and the sizes of the transmit and receive apertures ($d_{\tt T}$ and $d_{\tt R}$, respectively) are prime candidates for upward adjustment. Based on the required minimum SNR, these assumptions will further increase the estimate (\ref{eq:n-lens-dura}), helping to develop meaningful search strategy.  

Note that the maximum brightness would only be observed if the impact parameter is very small, and even then, only briefly. In that time, the lens will brighten featuring the amplified laser signal in accord with (\ref{eq:powed-l-Q}), assuming signal-dominated transmission. The spatial structure of the received beam is the beam width of  $d_{\tt beam}$ from (\ref{eq:rE}), which is crossed by Earth in just
{}
\begin{eqnarray}
\tau_{\tt beam}=\frac{d_{\tt beam}}{v_\oplus}\simeq 0.11\,\Big(\frac{z}{10\,{\rm pc}}\Big) \Big(\frac{650\,{\rm AU}}{z_0}\Big)\Big(\frac{d_{\tt T}}{1~{\rm m}}\Big)\Big(\frac{30~{\rm km/s}}{v_\oplus}\Big)~{\rm s}.
 \label{eq:size}
\end{eqnarray}
The region $d_{\tt beam}$ where the lensing reaches its maximal value of (\ref{eq:powed-l-Q}) and detection sensitivity (\ref{eq:powed-l-Qs}).  This is a short but very bright event that can be detected with current technology, even if receiver is not exactly on the optical axis.

What can be observed a photometric campaign is the varying light amplification as the Earth moves relative to direction to the lensing star with angular separation between them in units of the projected beam spot, $u={\theta}/{\theta_{\tt beam}}$, with $\theta_{\tt beam}=d_{\tt beam}/z$, which may be expressed as 
{}
\begin{equation}
u(t,t_0,u_0,\Delta t_{\tt beam})=\sqrt{u_0^2+(t-t_0)^2/\Delta t^2_{\tt beam}},
\label{eq:u}
\end{equation}
where $t_0$ is the time of closest alignment and $u_0$ is the impact parameter of the event, i.e., the angular separation of the source from the lens at $t_0$ expressed  in units of $\theta_{\tt beam}$. The quantity $\Delta t_{\tt beam}$ represents the characteristic time scale of the event and is given by (\ref{eq:n-lens-dura}). The objective is to monitor the brightness of the nearby stars for a signature of a microlensing event. The relevant light  amplification factor is characteristic for the lensing within the weak interference region of the lens  \citep{Schneider-Ehlers-Falco:1992,Gaudi:2012,Turyshev-Toth:2023}, which is given as below:
{}
  \begin{eqnarray}
A_{\tt weak.int}= \frac{u^2+2}{u\sqrt{u^2+4}}.
    \label{eq:a12_amp}
\end{eqnarray}

What is actually observed is the change in the flux of the transmitting lens given as
{}
\begin{eqnarray}
\Delta F(t,t_0,u_0,\Delta t_{\tt beam})=\Big(A_{\tt weak.int}(t,t_0,u_0,\Delta t_{\tt beam})-1\Big)F_0,
  \label{eq:Amp10}
\end{eqnarray}
where $A_{\tt weak.int}$ is from (\ref{eq:a12_amp}) and $F_0$ is the nominal flux of the nearby star. In the transmission scenario considered above, the flux $\Delta F$ changes twice a year and is correlated with the Earth's orbital motion. 

\subsection{Plausible detection strategies}
\label{sec:det-stra}

To optimize observations of such transient  events,  a cooperative telescope network is essential. This network would scan the sky to detect repeatable microlensing events originating at nearby stars with no stellar background that could be responsible.
Integrating multiple facilities could enhance spatial resolution, crucial for confident detection and increasing the probability of capturing transmitted signals. Any single telescope in the network can perform a broad spectrum of measurements. As a result, a synchronized network of telescopes will be capable of generating detailed imagery of the beam, conducting thorough analyses of its brightness variability.

Although resolving the Einstein rings with the current generation of optical instruments will be challenging, a collaborative network of  astronomical facilities may yield valuable information.   
With each facility recording brightness measurements and providing its heliocentric position information as a function time, one can use analytical tools available to monitor and study properties of the microlensing event as it unfolds.  We envision using both large telescopes and constellations of small telescopes to capture the photometric data produced by lensing events using the already  established photonics and optical engineering technologies. These systems will work in tandem, mutually supplementing observations and offering vital data for predicting events. Several telescopes either in space or both in space and on the ground can be combined in a network to monitor the star's intensity as lensing events unfold \cite{Turyshev-Toth:2023}. 

Generally, the search strategy can be formulated using expression (\ref{eq:power_rfin}). By converting this into a formula for photon flux, expressed as $Q_{\tt R} = (\lambda/hc)P_{\tt R}$, we obtain the following result:
{}
\begin{eqnarray}
Q_{\tt R}(\rho_0)&=&\Big(\frac{\lambda}{hc}\Big)P_{\tt T}\frac{\pi ({\textstyle\frac{1}{2}}d_{\tt R})^2}{\pi (z_0+z)^2}\Big(\frac{d_{\tt T}}{\lambda}\Big)^2\,\frac{4\sqrt{2r_g \tilde z}}{d_{\tt T}}\,\mu(\rho_0).
\label{eq:p-flux}
\end{eqnarray}
It is convenient to separate the terms in this expression that relate to the transmitter, the lens and the receiver.  For that, for $z\gg z_0$, we approximate the impact parameter as $b=\sqrt{2r_g \tilde z}\simeq R_{\tt L}$, equivalently transforming (\ref{eq:p-flux}) as below 
{}
\begin{eqnarray}
Q_{\tt R}(\rho_0)&\simeq &
\Big[\frac{P_{\tt T}d_{\tt T}}{hc}\Big]_{\rm transmitter} \Big[\frac{R_{\tt L}}{z^2}\Big]_{\rm lens}\Big[\frac{d_{\tt R}^2}{\lambda}\Big]_{\rm receiver}\mu(\rho_0).
\label{eq:p-flux2}
\end{eqnarray}

Clearly, the lens parameters $R_{\tt L}$ and $z$ are known for each target star. Assuming a specific telescope, $d_{\tt R}$ is fixed. With this information, one can conduct a search for a signal $Q_{\tt R} = Q_{\tt R}(\lambda, P_{\tt T}d_{\tt T})$ for a particular transmission wavelength $\lambda$ by varying the combination of the transmitter parameters $P_{\tt T}d_{\tt T}$. Consequently, for each star and a given  astronomical facility, for  each specific laser wavelength, the search space can be conceptually represented as $Q_{\tt R} = Q_{\tt R}\big(\lambda; P_{\tt T}d_{\tt T}\big)$. This approach could be used to develop a robust search strategy relying on the capabilities of current technology.    

We note that directing an interstellar signal at a specific point in Earth's heliocentric orbit could lead to false negatives. In such instances, a persistent signal, correlating with Earth's orbital period, might be observed annually. If this signal is sufficiently weak, it could be mistaken for background noise and disregarded. Therefore, it is critical to observe the targets across multiple spectral bands. Detecting periodic anomalous photometric behavior in one or more channels—especially if multiple laser wavelengths are employed but not uniformly—could signal the presence of the  signal of interest. Furthermore, a network of spatially distributed telescopes could facilitate early alerts and enable independent measurements, especially as they will sample  the data with different values of the factor $\mu(\rho_0)$.

Once the signal is detected, the spatial distribution of receivers is invaluable, as each will capture a distinct dataset by traveling through the signal along a different path, as given by $\mu(\rho_0)$. Correlating the photometric and spectral data from each path enables the reconstruction of the beam's full profile as it projected onto the solar system. Integrating this information with spectral data from multiple channels reveals the transmitter's specific features encoded in the beam, such as its power, shape, design, and propulsion capabilities. Additionally, if the optical signal contains encoded information, transmitted via a set of particular patterns, this information will become accessible as well.

As a result, coordinating ground- and space-based telescope observations is important to detect the  structure of the PSF of the transmitting lens as the receiver traverses the projected light field. To do that, evaluating the capabilities of current and upcoming astronomical facilities (see \cite{Turyshev-Toth:2023} for details) including the James Webb Space Telescope, the Nancy Grace Roman Space Telescope, Euclid, the Vera C. Rubin Observatory's LSST, the Thirty Meter Telescope, and the Extremely Large Telescope, complemented by a network of smaller telescopes is timely, as these facilities may be able to facilitate the search for the faint transient signals transmitted to us by our galactic neighbors.

Although we considered the case where transmitter points at a spot at 1\,AU from the Sun, other transmission scenarios exist, including one that involves directly targeting Earth and compensating for its orbital motion \cite{Howard-etal:2004}. This particular approach would necessitate enhanced propulsion and targeting capabilities for the transmitting spacecraft, thereby adding complexity to the design of both the spacecraft and its mission. Additionally, in such a scenario, the lensing star would display additional brightness due to persistent lensing effect, further complicating detection efforts if only a single telescope is employed. However, a coordinated network of ground- and space-based telescopes, discussed here, will be capable of effectively manage this transmission scenario thus improving chances of signal detection. Furthermore, monitoring the target star across various spectral channels and from diverse spatial positions enhances the chances of detecting unusual photometric variability that might manifest in one or several narrow spectral bands.

\section{Discussion} 
\label{sec:summary}

Every star is destined to form an Einstein ring.  This ring may form naturally with distant sources present in the background (see relevant discussion in \cite{Turyshev-Toth:2023}) or through advanced technology, deliberately deployed to send us a message, the case discussed in \cite{Turyshev:2024} and considered in this paper. The position of an Einstein ring around a star, as detailed in (\ref{eq:theta1-t}), depends on our distance to the star, the star's mass, and the transmitter's location. Although we can not resolve the rings that may form around nearby stellar lenses by light transmitted to us from their focal regions, with this information, we can initiate our search for the laser transmissions enabled by gravitational lensing. 

Through a synergistic effort, we will soon travel to the focal region of the solar gravitational lens (SGL), enabling direct high-resolution imaging and spectroscopy of exoplanets in our galactic neighborhood \cite{Helvajian-etal:2023,Friedman-etal:2024,Turyshev-etal:2023}.
However, before this, we must also survey nearby stars for potential signs of signals sent through gravitational lensing-enabled transmission links. Upon the mission's arrival at the focal region of the SGL, we will have the capacity to observe the surface of exoplanets and initiate meaningful information exchanges.

In the early phases of this process, a network of astronomical ground- and space-based astronomical facilities may provide valuable spatial, spectral and temporal information as microlensing events unfold.  This data may be processed with existing tools \cite{Turyshev-Toth:2023} to recover the properties of the signal and potentially the information encoded within.

To advance such a  moment, we considered the propagation of EM waves in the vicinity of a monopole gravitational lens. We explored a scenario with near-perfect alignment among the transmitter, the lens, and the receiver. In this axially-symmetric case, we successfully solved the relevant diffraction integrals analytically, yielding valuable insights.

We explored the case where a transmitter is in the focal region of a nearby star, with the receiver positioned within our solar system. Although one must account for other light sources near the lens, their impact on the achievable SNRs is expected to be minimal. Given the high sensitivities, especially in the well-motivated signal-dominated transmission regime, we considered the strategies in the search for transmitted signals using existing astronomical facilities or involving development of new collaborative spatially dispersed  facilities dedicated for this purpose.  

Our findings reveal that microlensing events, corresponding to laser signals transmitted through nearby gravitational lenses, are transient, requiring good and deliberate pointing on the part of a transmitter but notably bright and spatially widened, making them detectable with the current generation of instruments. Given their brightness, these events can be detected and analyzed through a dedicated campaign of photometric and spectroscopic observations. Deploying a spatially disperse network of collaborative astronomical facilities can enhance detection sensitivity. It may also be used to learn about the transmitter and to determine if any message was transmitted.  
As a result, we have demonstrated the feasibility of establishing interstellar power transmission links via gravitational lensing, while also confirming our technological readiness to receive such signals. It's time to develop and launch a search campaign.

Clearly, more work is still required, including: i) A comprehensive analysis of the propagation of annular Gaussian laser beams in the vicinity of gravitational lenses, as was done in \cite{Turyshev-Toth:2017} for plane waves. ii) A detailed exploration of the caustic patterns formed in the focal region of a gravitational lens by incident annular beams, similarly to \cite{Turyshev-Toth:2021-caustics}. iii) Consider spectral signals available with gravitational lensing, as in \cite{Turyshev-Toth:2022-mono_SNR}. iv) Numerical simulations to assess various transmission scenarios using incident Gaussian laser beams. v) Development of realistic SNR estimates using practical values for several key parameters, including detector quantum efficiency, optical throughput, spectroscopic sensitivity, instrumental noise, facility's location, among technical capabilities essential for the search. vi) Formulating requirements for reliable signal detection by conducting a comprehensive exploration of the available set of parameters. vii) A systematic analysis of the capabilities of existing facilities and those under construction to support the search efforts. Research in these areas is ongoing and results, when available, will be reported in subsequent publications. The outcome of this work could have profound implications for the ongoing search for interstellar transmissions. 

\begin{acknowledgments}
We acknowledge discussions with Jason T. Wright who kindly provided us with valuable comments and suggestions on various topics addressed in this document.
This work was performed at the Jet Propulsion Laboratory, California Institute of Technology, under a contract with the National Aeronautics and Space Administration. 

\end{acknowledgments}


\end{document}